\documentclass[conference,a4paper]{APSIPA2025}
\usepackage{amsmath}
\usepackage{graphicx}
\usepackage{multirow}
\usepackage{threeparttable}
\usepackage[backend=biber,style=ieee,]{biblatex}
\usepackage{amssymb,amsfonts}
\usepackage{booktabs}   
\usepackage{tabularx}   
\usepackage{array}      
\usepackage{textcomp}
\usepackage{xcolor}
\usepackage{hyperref}
\hypersetup{hidelinks=true}
\usepackage{subcaption}
\usepackage{flushend} 
\usepackage[normalem]{ulem}

\usepackage{geometry}
\usepackage{fancyhdr}

\fancypagestyle{firststyle}{
  \fancyhf{}
  \fancyhead[C]{2025 Asia Pacific Signal and Information Processing Association Annual Summit and Conference (APSIPA ASC)}
}

\begin{document}

\title{\textsc{Noro}: Noise-Robust One-shot Voice Conversion with Hidden Speaker Representation Learning}

\author{
\authorblockN{
Haorui He\authorrefmark{2}$^*$,
Yuchen Song\authorrefmark{2}$^*$,
Yuancheng Wang\authorrefmark{2}$^*$,
Haoyang Li\authorrefmark{3},
Xueyao Zhang\authorrefmark{2},
Li Wang\authorrefmark{2}, \\
Gongping Huang\authorrefmark{4}, 
Eng Siong Chng\authorrefmark{3}, and
Zhizheng Wu\authorrefmark{2}\thanks{Corresponding author: wuzhizheng@cuhk.edu.cn}
}

\authorblockA{
\authorrefmark{2}
School of Data Science, Chinese University of Hong Kong, Shenzhen, 518172, China}

\authorblockA{
\authorrefmark{3}
School of Computer Science and Engineering, Nanyang Technological University, Singapore 639798, Singapore}

\authorblockA{
\authorrefmark{4}
School of Electronic Information, Wuhan University, Wuhan 430072, China}
}

\maketitle
\thispagestyle{firststyle}
\pagestyle{fancy}

\begingroup
\renewcommand{\thefootnote}{\fnsymbol{footnote}} 
\footnotetext[1]{Equal contribution. Listed in alphabetical order.}
\endgroup

\begin{abstract}
The effectiveness of one-shot voice conversion (VC) decreases in real-world scenarios where reference speeches, which are often sourced from the internet, contain various disturbances like background noise. To address this issue, we introduce Noro, a noise-robust one-shot VC system. Noro features innovative components tailored for VC using noisy reference speeches, including a dual-branch reference encoding module and a noise-agnostic contrastive speaker loss. Experimental results demonstrate that Noro outperforms our baseline system in both clean and noisy scenarios, highlighting its efficacy for real-world applications. Additionally, we investigate the hidden speaker representation capabilities of our baseline system by repurposing its reference encoder as a speaker encoder. The results show that it is competitive with several advanced self-supervised learning models for speaker representation under the SUPERB settings, highlighting the potential for advancing speaker representation learning through one-shot VC tasks.
\end{abstract}


\section{Introduction}
One-shot voice conversion (VC) alters the timbre of speech from a source speaker to that of a target speaker using just one reference speech sample from the target speaker, while maintaining the semantic content of the original speech. This task has been extensively studied in the deep learning era, employing various methods and producing promising results~\cite{diffvc,SEF-VC,amphion,emilia,emilialarge}.
However, the strong performance of these one-shot VC methods is mainly demonstrated in controlled academic settings, where high-quality data is used for both training and evaluation. As noted in~\cite{robustvc-how-far,robustvc,noisy-to-noisy,flowvc}, their effectiveness significantly decreases in more challenging real-world scenarios, where reference speeches often collected from the internet are characterized by various interferences, such as background noise, leading to suboptimal converted speech. 

Few works have addressed this challenge to achieve noise-robust VC. Some researchers~\cite{vc_se_1, vc_se_2} attempt to integrate a pre-trained or jointly trained speech enhancement module into standard VC systems to improve robustness. However, these methods are not end-to-end and inevitably increase computational costs during inference. Other researchers~\cite{robustvc, robustvc-adv, flowvc,igarashi2024noise} suggest training strategies, such as data augmentation~\cite{robustvc}, to mitigate the impact of noise. However, there is still room for improvement, especially in extremely noisy conditions where the Signal-to-Noise Ratio (SNR) is below 5 dB~\cite{robustvc-adv}.

To address these challenges, we introduce a novel noise-robust one-shot VC system named Noro. Noro is built on a one-shot VC baseline system based on diffusion. This model is trained to generate speech from pitch and semantic representations obtained through a source encoder and speaker timbre representations via a reference encoder. To achieve noise-robustness with noisy reference speeches, we focus on learning speaker timbre representations that are agnostic to noise. We hypothesize that maintaining noise-agnostic input representations for the acoustic model will prevent its performance from deteriorating due to noise interference.

Therefore, we propose the following designs:
Firstly, we introduce a dual-branch reference encoding module. This module consists of two transformer encoders with shared weights: one branch encodes clean reference speech, while the other encodes its noisy counterpart generated through data augmentation.
Next, we devise a noise-agnostic contrastive speaker loss to maximize the similarity between samples (whether clean or noisy) from the same speaker while minimizing it for those from different speakers. This loss ensures that the dual-branch reference encoding module learns to represent speaker timbre independent of the acoustic environment of the reference speeches.
Experimental results show that Noro significantly enhances the robustness of our baseline system in diverse noisy environments, with only a minor reduction in performance in clean settings, making it well-suited for real-world applications involving noisy reference speeches.

Additionally, inspired by~\cite{tts4sv,tts4sv2}, we further investigate the speaker representation capabilities of our baseline system. The motivation stems from the fact that the reference encoder within the VC systems, trained to encode the vocal timbre of various speakers, including unseen ones, for one-shot VC, inherently functions as a speaker encoder. This encoder develops its capability to represent speakers in a self-supervised learning (SSL) manner without needing explicit speaker labels. Thus, we argue that the reference encoder might possess hidden abilities to function effectively as an SSL speaker encoder.
To validate this hypothesis, we employ a pre-trained reference encoder from our baseline one-shot VC system as a speaker encoder, denoted as VC-SPK2VEC, and assess its speaker representation ability through a speaker verification task under the SUPERB~\cite{SUPERB} setting. We compare its performance against several state-of-the-art (SOTA) SSL models, including Wav2vec~\cite{wav2vec}, Wav2vec 2.0~\cite{wav2vec2}, HuBERT~\cite{Hubert}, and WavLM~\cite{wavlm}. Surprisingly, VC-SPK2VEC achieves a competitive Equal Error Rate (EER) of 5.32\%, comparable to these SOTA SSL models. This result confirms the effectiveness of VC-SPK2VEC as an SSL speaker encoder and underscores the potential for leveraging one-shot VC tasks in advancing speaker representation learning.

\section{Methodology}
\begin{figure}[t]
\centering
\includegraphics[width=\linewidth]{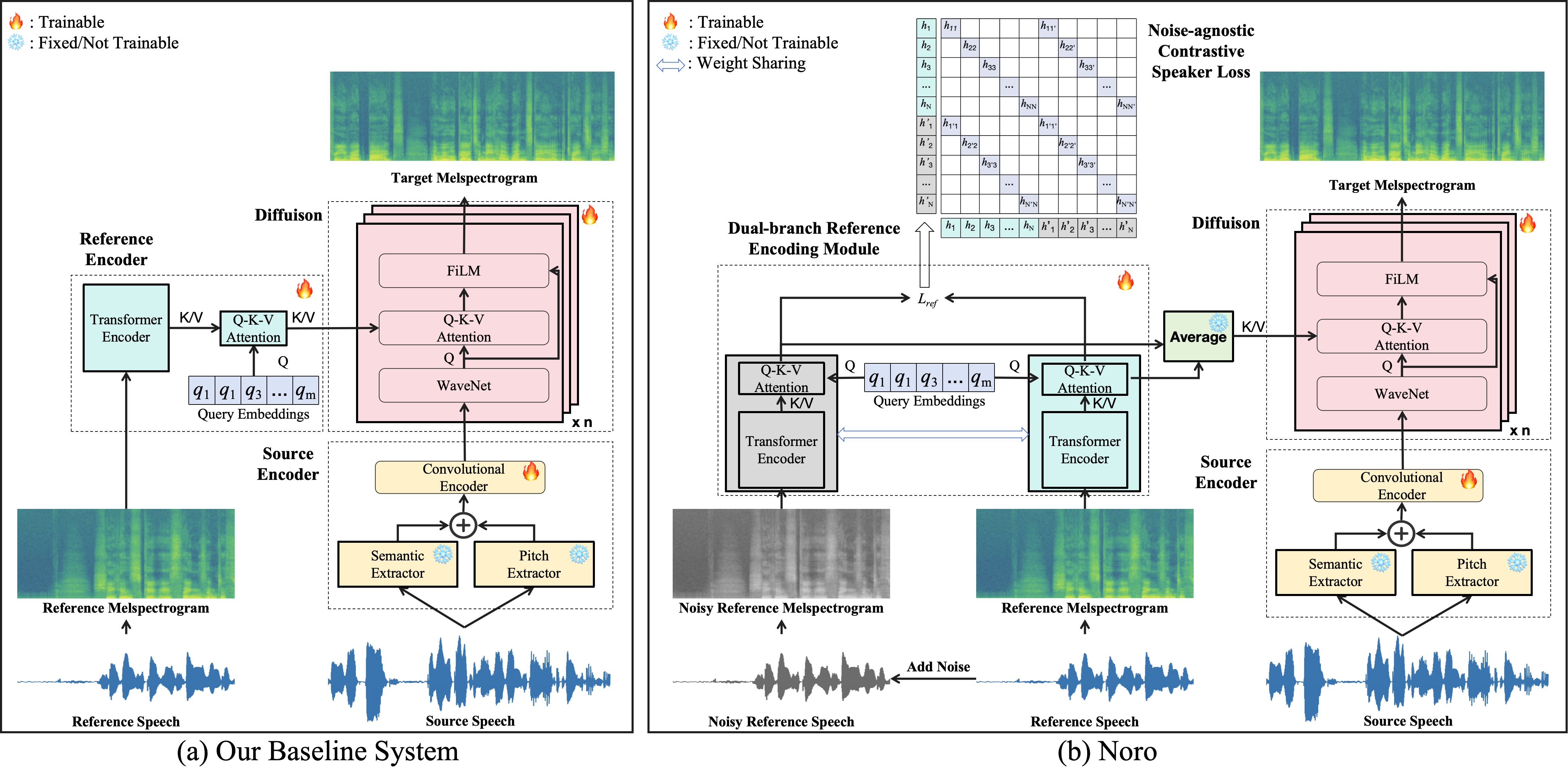}
\caption{Model architecture of our baseline system and Noro.}
\label{fig:noro} 
\end{figure}

In this section, we first introduce our baseline one-shot VC system and Noro. As depicted in Fig.~\ref{fig:noro}, Noro builds upon the baseline system, but replaces its original reference encoders with a dual-branch reference encoding module and a noise-agnostic contrastive speaker loss.

\subsection{The Design of Our Baseline System}
\subsubsection{Source Encoder}
\label{method-source}
To derive semantic representations from the source speech, our baseline system employs a pre-trained and frozen HuBERT model~\cite{Hubert, mhubert} as the semantic extractor. This model encodes the source speech into continuous embeddings. We then apply K-Means quantization to these embeddings. Instead of using K-Means cluster IDs (discrete semantic tokens), we use the quantized continuous embeddings, which preserve more information, for semantic representation. The resulting semantic representation for the source speech is denoted as $s_{\text{src}}$.
For pitch representations, we use the open-source software PyWorld to extract frame-level F0 values from the source speech. We apply min-max normalization to these values and use them for pitch representation, denoted as $p_{\text{src}}$.
After extracting the semantic and pitch representations, we employ a convolutional neural network (CNN)-based encoder to simultaneously encode $s_{\text{src}}$ and $p_{\text{src}}$, obtaining the final source representation $h_{\text{src}}$.

\subsubsection{Reference Encoder}
\label{method-reference}
To encode the speaker timbre in the reference speech, we use a cross-attention scheme similar to the speech prompting mechanism in NaturalSpeech 2~\cite{ns2}. Specifically, we first employ a Transformer-based reference encoder to convert the melspectrogram of the reference speech into a hidden sequence. Instead of allowing the diffusion model to directly attend to this hidden sequence, we employ two attention blocks. 
In the first attention block, we use $m$ randomly initialized query embeddings to attend to the hidden sequence, resulting in $h_{\text{ref}}$, a hidden sequence of length $m$ that represents the reference speaker's timbre at the utterance level. In the second attention block, the output hidden sequence of the WaveNet Layer serves as the query, while $h_{\text{ref}}$ functions as both the key and value. The attention results of the second attention block are then used as conditional information for a FiLM layer in the final acoustic modeling process.

\subsubsection{Diffusion Model}
\label{method-diffusion}
To enhance in-context learning for improved one-shot generation, each training sample is randomly segmented during the training phase. A segment comprising 25-45\% of the total sample is designated as the reference speech, while the remaining portion functions as both the source and target speech. We utilize a WaveNet-based diffusion model~\cite{ns2}, denoted as $s_\theta$, to predict the melspectrogram $\mathbf{z}_{\text{tgt}}$ of the target speech.
We describe both the diffusion (forward) process and the denoising (reverse) process using the stochastic differential equation. 
In the forward process, Gaussian noise is added to the melspectrogram $\mathbf{z}_{\text{tgt}}$:
$
\mathbf{z}_{\text{tgt}}^{t} = e^{-\frac{1}{2} \int_{0}^{t} \beta_s ds} \mathbf{z}_{\text{tgt}} + (\mathbf{I} - e^{- \int_{0}^{t} \beta_s ds}) \boldsymbol{\epsilon},
$
where $\beta_t$ represents the noise schedule function, $\boldsymbol{\epsilon}$ denotes randomly sampled Gaussian noise, and $t \in [0, 1]$.
The reverse process can be formulated as:
$
d \mathbf{z}_{\text{tgt}}^t = -\left(\frac{1}{2} \mathbf{z}_{\text{tgt}}^t + \nabla \log p_t(\mathbf{z}_{\text{tgt}}^t)\right) \beta_t \, dt + \sqrt{\beta_t} \boldsymbol{\epsilon}.
$

$s_\theta$ takes the noised melspectrogram $\mathbf{z}_{\text{tgt}}^t$, the time step $t$, and the conditions of the source representation $h_{\text{src}}$, containing both semantic and pitch information, and the reference representation $h_{\text{ref}}$, containing speaker timbre information, to estimate $\nabla \log p_t(\mathbf{z}_{\text{tgt}}^t)$. The loss function of the diffusion model can be written as:
$
\mathcal{L}_{\text{diff}} = \left\|s_\theta(\mathbf{z}_{\text{tgt}}^t, t, h_{\text{src}}, h_{\text{ref}}) - \nabla \log p_t(\mathbf{z}_{\text{tgt}}^t)\right\|_{1}.
$
During the inference stage, we gradually denoise $\mathbf{z}_{\text{tgt}}^t$ utilizing the estimated score $s_\theta(\mathbf{z}_{\text{tgt}}^t, t, s_{\text{src}}, p_{\text{src}}, h_{\text{ref}})$. This process initiates with Gaussian noise, initially sampled as $\mathbf{z}_{\text{tgt}}^1$.

\subsection{The Design of Noro}
\subsubsection{Dual-branch Reference Encoding Module} \label{method-ref-source}
To improve the noise robustness of the baseline system, we replaced its original reference encoder with a dual-branch reference encoding module. This module consists of two reference encoders sharing identical model weights as a dual-branch structure.
For each clean training sample, we randomly mix clean reference speech with eight types of noise from the DEMAND database~\cite{demand} at SNRs following a normal distribution (0,20) dB to create noisy reference speeches. Subsequently, the clean and noisy reference speeches are processed by separate reference encoders using the same query embeddings. This process generates clean and noisy reference representations $h_{\text{ref}}$ and $h'_{\text{ref}}$, respectively. 
Considering the weight-sharing scheme used in this dual-branch reference encoding module, the method described here qualifies as a ``training strategy" for enhancing our baseline system. Initially, we load the Noro system with weights from a pre-trained baseline system. During the training phase, the average of $h_{\text{ref}}$ and $h'_{\text{ref}}$ is fed into the diffusion model for acoustic modeling and acts as the keys and values in the second attention block. During the inference phase, only one reference encoder is utilized, and the Noro system's structure mirrors that of the baseline system.

\subsubsection{Noise-agnostic Contrastive Speaker Loss} 
After obtaining the representations of the clean speeches $h_{\text{ref}}$ and their noisy counterparts $h'_{\text{ref}}$, we introduce a noise-agnostic contrastive speaker loss. This loss function aims to maximize the similarity between samples (clean or noisy) from the same speaker while minimizing it for samples from different speakers.
First, as $h_{\text{ref}}$ and $h'_{\text{ref}}$ are hidden sequences of length $m$, we perform average pooling over the length dimension to form a comprehensive reference representation that captures the reference speaker's timbre at the utterance level. We then concatenate $h_{\text{ref}}$ and $h'_{\text{ref}}$ along the batch size dimension:
$
h_{\text{all}} = \left[ h_{\text{ref}}; h'_{\text{ref}} \right].
$
Given that the noisy speeches share the same speakers as the clean ones, $
y_{\text{all}} = \left[ y_{\text{spk}}; y_{\text{spk}} \right].
$
Finally, our proposed noise-agnostic contrastive speaker loss is formulated as:
$
\mathcal{L}_{\text{ref}} = \frac{1}{2N} \sum_{i=1}^{2N} \text{CrossEntropy} \left( \frac{h_{i} \cdot h_{j}^T}{\tau}, \mathbf{M}_{i,j} \right),
$
where $h_{i}$ and $h_{j}$ are the $i$-th and $j$-th reference representations in $h_{\text{all}}$, $\tau$ is the temperature parameter adjusting the scaling of the logits, $N$ is the batch size, and the mask matrix $\mathbf{M}_{i,j}$ is defined as follows:
$
\mathbf{M}_{i,j} =
\begin{cases} 
1 & \text{if } y_{i} = y_{j}, \\ 
0 & \text{otherwise},
\end{cases}
$
where $y_{i}$ and $y_{j}$ are the speaker labels for $h_{i}$ and $h_{j}$ respectively.
This loss ensures that the reference encoders represent the vocal timbre of different speakers regardless of noise interference, thus enhancing the robustness of the system.
The total loss function for the Noro system is defined as follows: $\mathcal{L}_{\text{total}} = \alpha \mathcal{L}_{\text{diff}} + \beta \mathcal{L}_{\text{ref}}$,
where \(\alpha\) and \(\beta\) are the weights assigned to each loss component.

\section{Experiments}

\subsection{One-shot VC Under Different Acoustic Environments}
\subsubsection{Experimental Setups}
To evaluate Noro, we use two acoustic settings: clean and noisy. In the clean setting, the high-quality VCTK Corpus, recorded in a studio, serves as the evaluation dataset. VCTK contains recordings from 110 English speakers with diverse accents, characterized by minimal noise interference. The test set comprises 150 randomly selected pairs of source and reference speech from VCTK. For the noisy setting, we introduce noise of unseen types from the training stage to the reference speeches in the test set at different SNR conditions to create challenging conditions to evaluate the robustness of VC systems.
Under these two settings, we compare Noro with our baseline system and three state-of-the-art one-shot VC models: FaCodec-VC \cite{ns3}, FreeVC \cite{freevc}, and DiffVC \cite{diffvc}. 
For both our systems, we employ the Libri-Light~\cite{librilight} dataset for training, which consists of 60k hours of 16 kHz speech data from over 8k distinct speakers. We leverage the official segmentation scripts to segment the speech data into approximately 15-second sequences by concatenating consecutive chunks with voice activity. Utterances longer than 30 seconds or shorter than 3 seconds are discarded. Detailed hyper-parameters is publicly available at Amphion.\footnote{\url{https://github.com/open-mmlab/Amphion/blob/main/egs/vc/Noro}}

We use both objective and subjective metrics to evaluate the systems.
For objective evaluation, we follow \cite{SEF-VC} to employ speaker embedding cosine similarity (SECS) and character error rate (CER). SECS is determined using a cutting-edge speaker representation model,\footnote{\url{https://huggingface.co/microsoft/wavlm-base-plus-sv}} with the results reflecting similarity to the original voice prompt. CER is calculated using an ASR model\footnote{\url{https://huggingface.co/facebook/hubert-large-ls960-ft}} to assess the intelligibility of the generated speech.
For subjective evaluation, we follow \cite{ns2,ns3} to use the Comparative Mean Opinion Score (CMOS) and the Similarity Mean Opinion Score (SMOS) to evaluate naturalness and similarity, respectively. We use ten randomly selected pairs from the clean test set and another ten from the noisy test set. Twelve proficient English speakers conducted the assessments. The naturalness scores range from one (``Bad") to five (``Excellent"), and the similarity scores range from one (``Different speaker, sure") to four (``Same speaker, sure").

\subsubsection{Results and Analysis}
\begin{table}[ht]
    \centering
    \caption{The experimental results for objective evaluation.}
    \label{tab:vc}
    \resizebox{\linewidth}{!}{
    \begin{tabular}{lcccccc}
        \toprule
        \multirow{2}{*}{\textbf{Model}} & \multicolumn{2}{c}{\textbf{Clean}} & \multicolumn{2}{c}{\textbf{Noisy (0--5 dB)}} & \multicolumn{2}{c}{\textbf{Noisy (5--10 dB)}} \\
        \cmidrule(lr){2-3} \cmidrule(lr){4-5} \cmidrule(lr){6-7}
        & \textbf{CER $\downarrow$} & \textbf{SECS $\uparrow$} & \textbf{CER $\downarrow$} & \textbf{SECS $\uparrow$} & \textbf{CER $\downarrow$} & \textbf{SECS $\uparrow$}  \\
        \midrule
        Ground Truth & 1.37 & 90.4 & 1.37 & 90.4 & 1.37 & 90.4\\ 
        \midrule 
        FaCodec-VC & 3.66 & 82.33 & 3.56 & 74.78 & 3.26 & 77.88  \\
        FreeVC & 3.31 & 80.07 & 3.49 & 73.83 & 3.13 & 76.14 \\
        DiffVC & 16.91 & 85.32 & 15.72 & 82.00 & 13.91 & 82.33  \\
        \midrule 
        Our Baseline & 4.71 & 82.35 & 7.26 & 77.28 & 6.43 & 78.89   \\
        Noro & 4.74 & 82.38 & 4.66 & 80.09 & 5.06 & 80.71  \\
        \bottomrule
    \end{tabular}}
\end{table}

\begin{table}[ht]
    \centering
    \caption{The experimental results for subjective evaluation.}
    \label{tab:vc-sub}
    \resizebox{\linewidth}{!}{%
    \begin{tabular}{cccccccc}
        \toprule
        \textbf{Model} & \multicolumn{2}{c}{\textbf{Clean}} & \multicolumn{2}{c}{\textbf{Noisy (0--5 dB)}}\\
        \cmidrule(lr){2-3} \cmidrule(lr){4-5}
         & \textbf{CMOS $\uparrow$} & \textbf{SMOS $\uparrow$} & \textbf{CMOS $\uparrow$} & \textbf{SMOS $\uparrow$}\\
        \midrule
        Ground Truth & 4.04 & 3.90 & 4.04 & 3.90\\
        \midrule
        Our Baseline & \textbf{3.29} & \textbf{3.02} & 2.09 & 2.75\\
        Noro & 3.16 & 2.90 & \textbf{2.95} & \textbf{2.97}\\
        \bottomrule
    \end{tabular}}
\end{table}

Table~\ref{tab:vc} presents the objective evaluation results for the Noro and baseline systems in both clean and noisy environments. In clean conditions, the baseline system achieved a CER of 4.71 and a SECS of 82.35, while Noro had a comparable CER of 4.74 and a SECS of 82.38. Both systems demonstrated effective one-shot voice conversion, accurately preserving semantic content and adapting to speaker timbre.
Under noisy conditions, the baseline system's performance deteriorated significantly, with the CER increasing to 7.26 and the SECS dropping to 77.28 at 0--5 dB, reflecting its poor noise-handling capabilities. In contrast, Noro exhibited significantly better noise robustness, with a CER of 4.66 and a SECS of 80.09, indicating its reliability in real-world, noisy environments.
The subjective evaluation results in Table~\ref{tab:vc-sub} are consistent with the objective findings. Both systems performed well in clean conditions, maintaining speech quality and speaker identity. However, under noisy conditions, the baseline system’s performance declined sharply, with a CMOS of 2.09 and an SMOS of 2.75, while Noro maintained significantly higher scores, with a CMOS of 2.95 and an SMOS of 2.97. These results underscore Noro’s superior noise robustness and its practicality for use in diverse noisy environments.

\begin{figure}[tbbp]
    \centering
    \subfloat[Our Baseline System]{\includegraphics[width=.47\columnwidth]{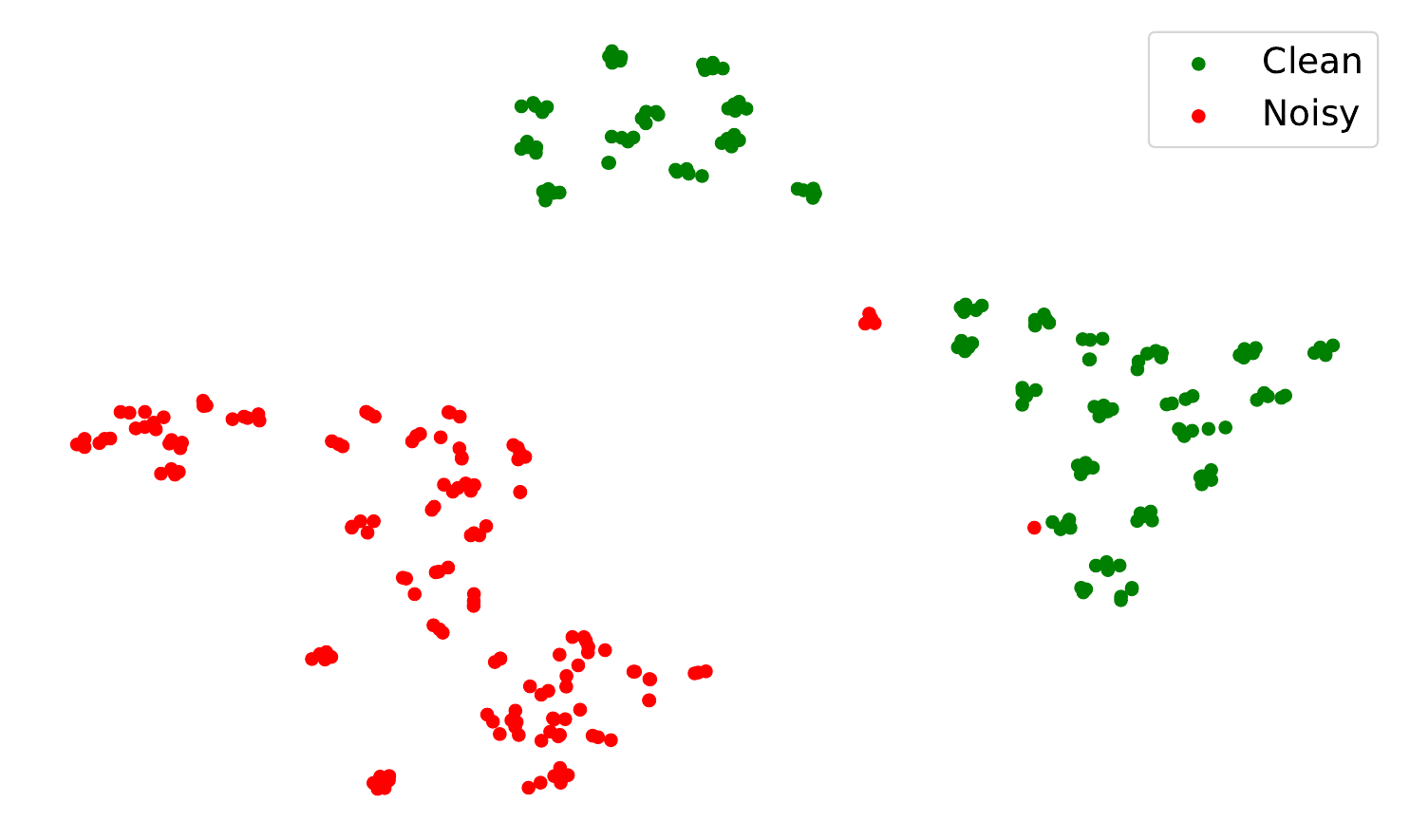} \label{fig:clean_embeddings}} \hspace{2pt}
    \subfloat[Noro]{\includegraphics[width=.47\columnwidth]{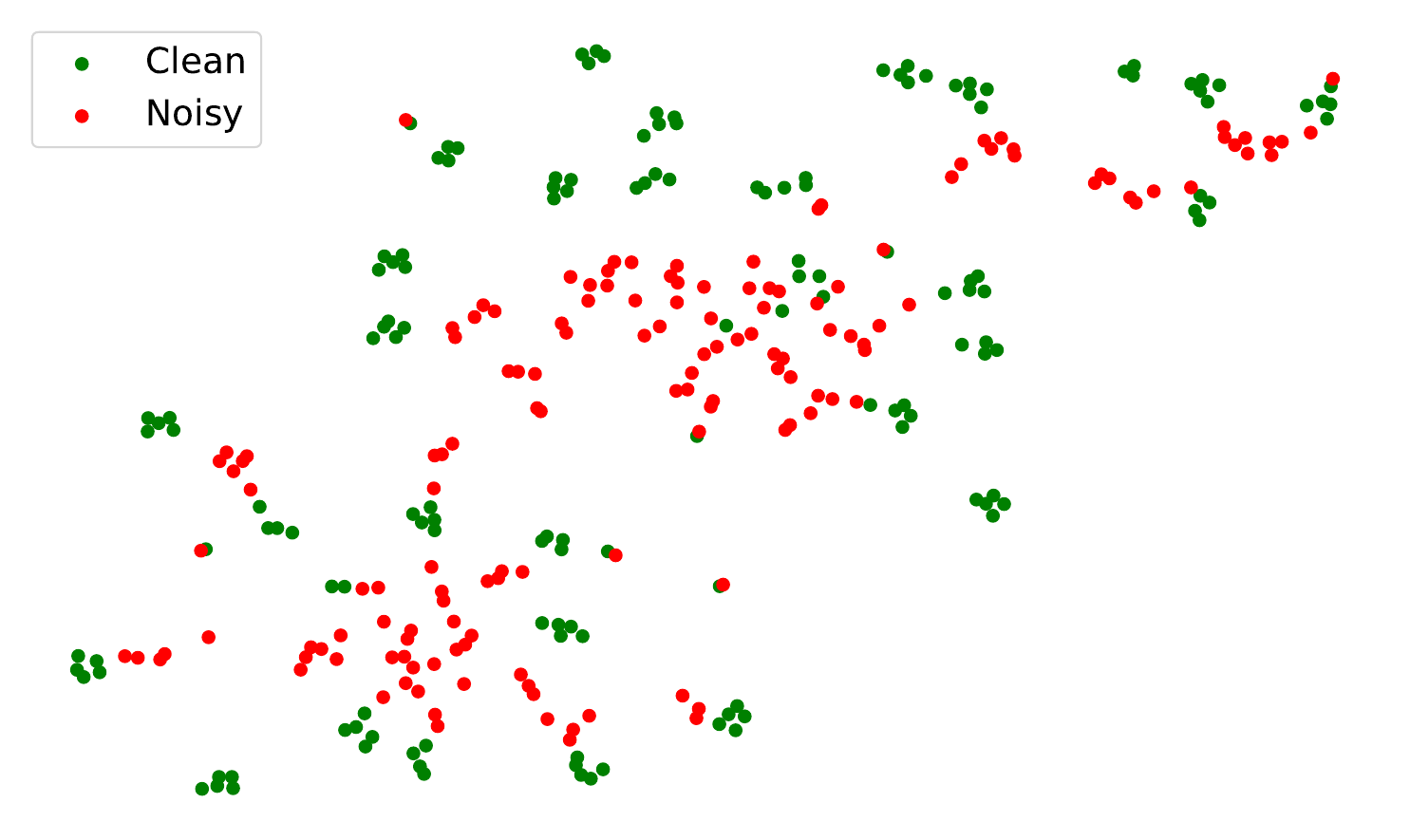} \label{fig:clean_noisy}}
    \caption{t-SNE visualization of reference representations.}
    \label{fig:emb}
\end{figure}

We further visualize the reference representation $h_{\text{ref}}$ produced by the baseline system and Noro using t-SNE, as shown in Fig.~\ref{fig:emb}. In the baseline system, the reference representations for clean and noisy speech show a clear separation, which contributes to the significant drop in performance when the reference speech is noisy. In contrast, Noro’s reference representations for both clean and noisy speech are well-mixed. This is due to Noro’s use of the proposed dual-branch reference encoding module and noise-agnostic contrastive speaker loss during training, which enables it to learn a noise-agnostic reference representation. This capability accounts for Noro’s superior performance in diverse noisy environments.

\subsection{One-shot VC as SSL for Speaker Representation}
\subsubsection{Experimental Setups}
The reference encoder effectively captures speaker timbre characteristics. Here, we adapt the reference encoder—originally designed for one-shot VC—into a self-supervised learning (SSL) speaker encoder, named VC-SPK2VEC. We evaluate its performance in speaker representation by comparing it with several widely used pre-trained SSL models \cite{CPC,wav2vec,vq-wav2vec,wav2vec2,Hubert,wavlm} under the SUPERB framework.
Following the SUPERB~\cite{SUPERB} setup, we employ SSL features for speaker verification. Input waveforms are processed through a frozen pre-trained SSL model to extract frame-level features, and a simple x-vector model serves as the downstream model, designed to classify whether two utterances belong to the same speaker, framing this as a binary classification task.
All models are fine-tuned using the VoxCeleb2 dataset, augmented with noise, and tested on VoxCeleb1~\cite{voxceleb}. For traditional SSL models, the final speaker representation is produced by computing the weighted sum of hidden states from each layer: $\hat{\mathbf{o}}_t = \sum_{l=1}^{L} w_l \cdot h_{l}^{t}$, where $h_{l}^{t}$ represents the hidden state from layer $l$ at time $t$, and $w_l$ denotes the normalized weight for each layer. In contrast, for VC-SPK2VEC, the final representation is taken from the hidden states of the reference encoder’s transformer encoder’s last layer. The x-vector downstream model utilizes an additive-margin softmax loss, maintaining the same hyperparameters as outlined in~\cite{voxceleb}. Cosine similarity is then used to compute the matching scores between speaker embeddings for each enrollment-test utterance pair.

\subsubsection{Results and Analysis}

\begin{table}[t]
\caption{The SV performance comparison. The results for these baseline models are sourced from~\cite{SUPERB} and~\cite{wavlm}.}
\label{speaker}
\centering
\resizebox{\linewidth}{!}{%
\begin{tabular}{l c c c}
\hline
Model & Params. & Data & EER (\%) \\
\hline
FBANK & - & - & 9.56 \\
\hline
Modified CPC~\cite{CPC} & 1.84M & 60k hr & 12.86 \\
wav2vec~\cite{wav2vec} & 32.54M & 960 hr & 7.99 \\
vq-wav2vec~\cite{vq-wav2vec} & 34.15M & 960 hr & 10.38 \\
wav2vec 2.0 Base~\cite{wav2vec2} & 95.04M & 960 hr & 6.02 \\
HuBERT Base~\cite{Hubert} & 94.68M & 960 hr & 5.11 \\
WavLM Base~\cite{wavlm} & 94.70M & 960 hr & 4.69 \\
WavLM Base+~\cite{wavlm} & 94.70M & 94k hr & 4.07 \\
\hline
wav2vec 2.0 Large~\cite{wav2vec2} & 317.38M & 60k hr & 5.65 \\
HuBERT Large~\cite{Hubert} & 316.61M & 60k hr & 5.98 \\
WavLM Large~\cite{wavlm} & 316.62M & 94k hr & \textbf{3.77} \\
\hline
VC-SPK2VEC & 72.4M & 60k hr & 5.32 \\
\hline
\end{tabular}
}
\end{table}

Table~\ref{speaker} compares VC-SPK2VEC with various baseline models on the speaker verification (SV) task. VC-SPK2VEC achieves a competitive equal error rate (EER) of 5.32\%. While it trails slightly behind HuBERT Base (5.11\%) and the three WavLM model variants (4.69\%, 4.07\%, and 3.77\% for Base, Base+, and Large, respectively), it outperforms other baseline models such as wav2vec 2.0 Base, wav2vec 2.0 Large, and HuBERT Large. Notably, VC-SPK2VEC uses a comparatively smaller model size of 72.4M parameters. These results underscore VC-SPK2VEC's effectiveness as a self-supervised speaker encoder and highlight its comparable performance to SOTA SSL models.
These findings highlight the great potential of leveraging speaker representation learning in one-shot speech generation tasks, as initially explored in~\cite{tts4sv,tts4sv2}. With the advent of speech generation models trained on extensive datasets exceeding 100k hours of speech data, unifying speaker representation learning with generative speech tasks like voice conversion (VC) may lead to further advancements in both fields. Future research could investigate the relationship between the capability of speaker representation and factors such as the number of speakers included in the training set of one-shot VC systems or the model size of the reference encoder in these systems.

\section{Conclusions}
In this paper, we introduced Noro, a noise-robust one-shot voice conversion system with a dual-branch reference encoding module and noise-agnostic contrastive speaker loss to improve performance in diverse noisy environments. Our experiments show that Noro significantly outperforms the baseline in noisy settings with minimal performance loss in clean conditions, making it a practical solution for real-world use.  
We also repurposed the reference encoder as an SSL speaker encoder, VC-SPK2VEC, and demonstrated its competitive performance on the SUPERB benchmark compared to advanced SSL models. This suggests a promising direction for unifying speaker representation learning and voice conversion, with potential for leveraging large-scale voice conversion systems and unlabeled data for further advancements in both fields.

\newpage
\section*{Acknowledgment}
The work is supported by the National Natural Science Foundation of China (No. 62376237), the 2023 Shenzhen Stability Science Program, the Shenzhen Science and Technology Program (No. ZDSYS20230626091302006), and the Program for Guangdong Introducing Innovative and Entrepreneurial Teams (No. 2023ZT10X044).
\printbibliography

@inproceedings{emilialarge,
    author={He, Haorui and Shang, Zengqiang and Wang, Chaoren and Li, Xuyuan and Gu, Yicheng and Hua, Hua and Liu, Liwei and Yang, Chen and Li, Jiaqi and Shi, Peiyang and Wang, Yuancheng and Chen, Kai and Zhang, Pengyuan and Wu, Zhizheng},
    title={{Emilia: A Large-Scale, Extensive, Multilingual, and Diverse Dataset for Speech Generation}},
    booktitle={arXiv:2501.15907},
    year={2025}
}

@article{wavlm,
  title={{Wavlm: Large-scale Self-supervised Pre-training for Full Stack Speech Processing}},
  author={Chen, Sanyuan and Wang, Chengyi and Chen, Zhengyang and Wu, Yu and Liu, Shujie and Chen, Zhuo and Li, Jinyu and Kanda, Naoyuki and Yoshioka, Takuya and Xiao, Xiong and others},
  journal={IEEE Journal of Selected Topics in Signal Processing},
  volume={16},
  number={6},
  pages={1505--1518},
  year={2022},
}

@inproceedings{SUPERB,
  title={{SUPERB: Speech Processing Universal Performance Benchmark}},
  author={Yang, Shu-wen and Chi, Po-Han and Chuang, Yung-Sung and Lai, Cheng-I Jeff and Lakhotia, Kushal and Lin, Yist Y and Liu, Andy T and Shi, Jiatong and Chang, Xuankai and Lin, Guan-Ting and others},
  booktitle = {INTERSPEECH},
  year = {2021}
}

@inproceedings{tts4sv,
  title={{Learning Speaker Embedding from Text-to-Speech}},
  author={Cho, Jaejin and Zelasko, Piotr and Villalba, Jesús and Watanabe, Shinji and Dehak, Najim},
  booktitle={{INTERSPEECH}},
  year={2020}
}

@inproceedings{tts4sv2,
  title={{Improving Reconstruction Loss Based Speaker Embedding in Unsupervised and Semi-Supervised Scenarios}},
  author={Cho, Jaejin and Żelasko, Piotr and Villalba, Jesús and Dehak, Najim},
  booktitle={{ICASSP}},
  year={2021},
}

@article{voxceleb,
  title={{VoxCeleb: Large-Scale Speaker Verification in the Wild}},
  author={Nagrani, Arsha and Chung, Joon Son and Xie, Weidi and Zisserman, Andrew},
  journal={Computer Speech \& Language},
  volume={60},
  pages={101027},
  year={2020},
  publisher={Elsevier}
}

@inproceedings{CPC,
  author={Rivière, Morgane and Joulin, Armand and Mazaré, Pierre-Emmanuel and Dupoux, Emmanuel},
  booktitle={{ICASSP}}, 
  title={{Unsupervised Pretraining Transfers Well Across Languages}}, 
  year={2020},
}

@inproceedings{wav2vec,
  title={{Wav2vec: Unsupervised Pre-Training for Speech Recognition}},
  author={Schneider, Steffen and Baevski, Alexei and Collobert, Ronan and Auli, Michael},
  booktitle={{INTERSPEECH}},
  year={2019},
}

@inproceedings{vq-wav2vec,
  title={{Vq-Wav2vec: Self-Supervised Learning of Discrete Speech Representations}},
  author={Baevski, Alexei and Schneider, Steffen and Auli, Michael},
  booktitle={{ICLR}},
  year={2019}
}

@inproceedings{wav2vec2,
  title={{Wav2vec 2.0: A Framework for Self-Supervised Learning of Speech Representations}},
  author={Baevski, Alexei and Zhou, Yuhao and Mohamed, Abdelrahman and Auli, Michael},
   booktitle={{NeurIPS}},
  year={2020}
}

@article{Hubert,
  title={{Hubert: Self-Supervised Speech Representation Learning by Masked Prediction of Hidden Units}},
  author={Hsu, Wei-Ning and Bolte, Benjamin and Tsai, Yao-Hung Hubert and Lakhotia, Kushal and Salakhutdinov, Ruslan and Mohamed, Abdelrahman},
  journal={IEEE/ACM Transactions on Audio, Speech, and Language Processing},
  volume={29},
  pages={3451--3460},
  year={2021},
}

@article{ns3,
  title={{Naturalspeech 3: Zero-Shot Speech Synthesis with Factorized Codec and Diffusion Models}},
  author={Ju, Zeqian and Wang, Yuancheng and Shen, Kai and Tan, Xu and Xin, Detai and Yang, Dongchao and Liu, Yanqing and Leng, Yichong and Song, Kaitao and Tang, Siliang and others},
  journal={ICML},
  year={2024}
}

@inproceedings{ns2,
  title={{Naturalspeech 2: Latent Diffusion Models are Natural and Zero-Shot Speech and Singing Synthesizers}},
  author={Shen, Kai and Ju, Zeqian and Tan, Xu and Liu, Yanqing and Leng, Yichong and He, Lei and Qin, Tao and Zhao, Sheng and Bian, Jiang},
  booktitle={{ICLR}},
  year={2023}
}

@inproceedings{igarashi2024noise,
  title={{Noise-Robust Voice Conversion by Conditional Denoising Training Using Latent Variables of Recording Quality and Environment}},
  author={Igarashi, Takuto and Saito, Yuki and Seki, Kentaro and Takamichi, Shinnosuke and Yamamoto, Ryuichi and Tachibana, Kentaro and Saruwatari, Hiroshi},
  booktitle={{INTERSPEECH}},
  year={2024}
}

@inproceedings{emilia,
    author={He, Haorui and Shang, Zengqiang and Wang, Chaoren and Li, Xuyuan and Gu, Yicheng and Hua, Hua and Liu, Liwei and Yang, Chen and Li, Jiaqi and Shi, Peiyang and Wang, Yuancheng and Chen, Kai and Zhang, Pengyuan and Wu, Zhizheng},
    title={{Emilia: An Extensive, Multilingual, and Diverse Speech Dataset for Large-Scale Speech Generation}},
    booktitle={SLT},
    year={2024}
}

@inproceedings{amphion,
    author={Zhang, Xueyao and Xue, Liumeng and Gu, Yicheng and Wang, Yuancheng and Li, Jiaqi and He, Haorui and Wang, Chaoren and Song, Ting and Chen, Xi and Fang, Zihao and Chen, Haopeng and Zhang, Junan and Tang, Tze Ying and Zou, Lexiao and Wang, Mingxuan and Han, Jun and Chen, Kai and Li, Haizhou and Wu, Zhizheng},
    title={Amphion: An Open-Source Audio, Music and Speech Generation Toolkit},
    booktitle={SLT},
    year={2024}
}

@inproceedings{robustvc,
  title={{Toward Degradation-Robust Voice Conversion}},
  author={Huang, Chien-Yu and Chang, Kai-Wei and Lee, Hung-Yi},
  booktitle={{ICASSP}},
  year={2022},
}

@inproceedings{librilight,
  title={{Libri-Light: A Benchmark for ASR with Limited or No Supervision}},
  author={Kahn, Jacob and Riviere, Morgane and Zheng, Weiyi and Kharitonov, Evgeny and Xu, Qiantong and Mazaré, Pierre-Emmanuel and Karadayi, Julien and Liptchinsky, Vitaliy and Collobert, Ronan and Fuegen, Christian and others},
  booktitle={{ICASSP}},
  year={2020},
}

@inproceedings{demand,
  title={{Investigating RNN-Based Speech Enhancement Methods for Noise-Robust Text-to-Speech}},
  author={Valentini-Botinhao, Cassia and Wang, Xin and Takaki, Shinji and Yamagishi, Junichi},
  booktitle={{SSW}},
  year={2016}
}

@inproceedings{SEF-VC,
  title={{SEF-VC: Speaker Embedding Free Zero-Shot Voice Conversion with Cross Attention}},
  author={Li, Junjie and Guo, Yiwei and Chen, Xie and Yu, Kai},
  booktitle={{ICASSP}},
  year={2024},
}

@inproceedings{robustvc-how-far,
  title={{How Far are We from Robust Voice Conversion: A Survey}},
  author={Huang, Tzu-Hsien and Lin, Jheng-Hao and Lee, Hung-Yi},
  booktitle={{SLT}},
  year={2021},
}

@article{noisy-to-noisy,
  author={Xie, Chao and Toda, Tomoki},
  journal={IEEE/ACM Transactions on Audio, Speech, and Language Processing}, 
  title={{Noisy-to-Noisy Voice Conversion Under Variations of Noisy Condition}}, 
  year={2023},
  volume={31},
  pages={3871--3882},
}

@article{robustvc-adv,
  title={{Noise-Robust Voice Conversion with Domain Adversarial Training}},
  author={Du, Hongqiang and Xie, Lei and Li, Haizhou},
  journal={Neural Networks},
  volume={148},
  pages={74--84},
  year={2022},
}

@inproceedings{flowvc,
  title={{Learning Noise-Independent Speech Representation for High-Quality Voice Conversion for Noisy Target Speakers}},
  author={Xue, Liumeng and Yang, Shan and Hu, Na and Su, Dan and Xie, Lei},
  booktitle={{INTERSPEECH}},
  year={2021},
}

@inproceedings{vc_se_1,
  title={{Investigating RNN-Based Speech Enhancement Methods for Noise-Robust Text-to-Speech}},
  author={Valentini-Botinhao, Cassia and Wang, Xin and Takaki, Shinji and Yamagishi, Junichi},
  booktitle={{SSW}},
  year={2016}
}

@article{vc_se_2,
  title={{Speech Enhancement-Assisted Stargan Voice Conversion in Noisy Environments}},
  author={Chan, Yun-Ju and Peng, Chiang-Jen and Wang, Syu-Siang and Wang, Hsin-Min and Tsao, Yu and Chi, Tai-Shih},
  journal={arXiv preprint arXiv:2110.09923},
  year={2021}
}

@inproceedings{mhubert,
  title={{Textless Speech-to-Speech Translation on Real Data}},
  author={Lee, Ann and Gong, Hongyu and Duquenne, Paul-Ambroise and Schwenk, Holger and Chen, Peng-Jen and Wang, Changhan and Popuri, Sravya and Adi, Yossi and Pino, Juan and Gu, Jiatao and Hsu, Wei-Ning},
  booktitle={{NACCL}},
  year={2022},
}

@inproceedings{freevc,
  title={{FreeVC: Towards High-Quality Text-Free One-Shot Voice Conversion}},
  author={Li, Jingyi and Tu, Weiping and Xiao, Li},
  booktitle={{ICASSP}},
  year={2023},
}

@inproceedings{diffvc,
  title={{Diffusion-Based Voice Conversion with Fast Maximum Likelihood Sampling Scheme}},
  author={Popov, Vadim and Vovk, Ivan and Gogoryan, Vladimir and Sadekova, Tasnima and Kudinov, Mikhail Sergeevich and Wei, Jiansheng},
  booktitle={{ICLR}},
  year={2022}
}

\end{document}